\documentclass[12pt]{iopart}
\usepackage{graphicx}
\usepackage{cite}
\usepackage{hyperref}
\usepackage[utf8]{inputenc}
\usepackage[
top    = 2.75cm,
bottom = 2.50cm,
left   = 3.00cm,
right  = 2.50cm]{geometry}
\usepackage{fancyhdr}
\begin{document}

\title{Environmental instability of few-layer black phosphorus} 

\author{Joshua O. Island, Gary A. Steele, Herre S. J. van der Zant, Andres Castellanos-Gomez}
\ead{j.o.island@tudelft.nl, a.castellanosgomez@tudelft.nl}
\address{Kavli Institute of Nanoscience, Delft University of Technology, Lorentzweg 1, 2628 CJ Delft, The Netherlands}

\begin{abstract}
We study the environmental instability of mechanically exfoliated few-layer black phosphorus (BP). From continuous measurements of flake topography over several days, we observe an increase of over 200\% in volume due to the condensation of moisture from air. We find that long term exposure to ambient conditions results in a layer-by-layer etching process of BP flakes. Interestingly, flakes can be etched down to single layer (phosphorene) thicknesses. BP's strong affinity for water greatly modifies the performance of fabricated field-effect transistors (FETs) measured in ambient conditions. Upon exposure to air, we differentiate between two timescales for changes in BP FET transfer characteristics: a short timescale (minutes) in which a shift in the threshold voltage occurs due to physisorbed oxygen and nitrogen, and a long timescale (hours) in which strong p-type doping occurs from water absorption. Continuous measurements of BP FETs in air reveal eventual degradation and break-down of the channel material after several days due to the layer-by-layer etching process. 
\end{abstract}

\maketitle
The two-dimensional (2D), elemental semiconductor black phosphorus (BP) has ignited recent interest from the material science community \cite{ li14,liu14, gomez14, koenig14, xia14, liu142, kamalakar14, qiao14, das14, appalakondaiah12}. Unlike graphene, the prototype 2D material, BP presents a direct band-gap that is capable of manifesting large ON/OFF ratios and high mobilities when used as a channel material in a field effect transistor (FET) \cite{li14, koenig14}. In addition, the direct band-gap of BP varies with thickness from $\approx2$ eV in single layer, phosphorene, to $\approx0.3$ eV in bulk making it attractive for optoelectronic applications \cite{dai14, buscema14, engel14, tran14, low14, buscema142}. This is in contrast to the semiconducting transition metal dichalcogenides (TMDC) such as MoS$_2$ and WSe$_2$ which only present a direct band-gap in single layers \cite{mak10, zhao12, kumar12}. While the accolades for BP seem promising, a serious consideration that remains is degradation under ambient conditions \cite{gomez14, koenig14}. At its first synthesis \cite{bridgman14} and in subsequent studies \cite{nishii87}, hints have been made at the possibility of water absorption at the surface of BP but a thorough study on the environmental instability of BP is lacking. 

Here, we study the environmental instability of few-layer BP in ambient conditions. First, we characterize water absorption at the surface of thin, isolated flakes of BP by continuous atomic force microscopy (AFM) measurements of the topography over several days. We also study the effect of long term exposure to ambient conditions on the topography of the few-layer flakes and find that it yields a layer-by-layer etching process which can even reduce flakes to single layer thicknesses. The role of atmospheric adsorbates on the electrical properties are also studied by systematically comparing the field-effect characteristics of black phosphorus flakes in vacuum and air. We find that an initial exposure to air results in a shift of the threshold voltage to negative gate values and an overall decrease in the drain current of fabricated BP FETs. Longer exposure to air results in a strong p-type doping of the flakes which we attribute to the presence of absorbed water.

BP is the only layered, van der Waals material of the phosphorus allotropes and can be created by heating white phosphorus under high pressure \cite{bridgman14}. It has an orthorhombic crystal structure with phosphorus atoms covalently bonded to three neighboring atoms to form a puckered, single-layer honeycomb lattice \cite{brown65}. Single layers are bonded through van der Waals interaction which allows exfoliation of the bulk material into thin flakes much like graphite. Recent ab initio calculations reveal that the van der Waals interaction is mainly Keesom forces from out-of-plane dipole interaction \cite{du10}. Our further DFT calculations in a previous publication have shown that the out-of-plane dipole moment makes BP highly hydrophilic as it can interact with the dipole moment of water molecules \cite{gomez14}. In the simulations, water molecules are allowed to relax at the surface of BP and do so with their dipole moment pointing upwards with respect to the BP surface. This dipole interaction significantly distorts the BP crystal lattice which shrinks by 25\%. This suggests that water plays a significant role in the degradation of BP flakes.

Preparation of isolated few-layer BP flakes is carried out using a modified mechanical exfoliation technique outlined in Ref ~\citenum{gomez14}. Bulk, commercially available BP material (99.998\%, Smart Elements) is used in this study. Previous characterization by Raman and transmission electron microscopy (TEM) show that the material is highly crystalline and free from extended defects \cite{gomez14}. By employing a viscoelastic stamp (Gelfilm from Gelpak) as an intermediate substrate for exfoliation, thin BP flakes are transfered to a Si substrate (with 285 nm of SiO$_{2}$ capping layer) leaving less residue than when compared with using Scotch tape. Thin flakes are then identified by contrast under an optical microscope and subsequently chosen to track water absorption. We measured three flakes, of varying thicknesses, in detail over time. One is presented in the main text and the two others can be found in the Supplemental Information. 

Water condensation at the surface of few-layer BP is studied using an AFM. The AFM is operated in tapping mode to reduce tip-sample interaction and image droplet formation on the flake surface over time \cite{zhong93}. In Figure 1 we summarize the water absorption over a five day period for a BP flake ranging in an initial thickness from 8 nm at its thinnest part to 30 nm at its thickest part (average relative humidity was 60\% during measurements taken in ambient light conditions). A video clip made of the collected images is available in the Supplemental Information. Selected AFM scans over the total measurement period are shown in Figures 1(a)-(d). At 3 hours, Figure 1(a), water droplets (100 nm in diameter and 5 nm tall) have already formed at the flake surface and after five days, Figure 1(d), water completely covers the flake and results in a large convex meniscus. Line profiles taken at the same flake position for the scans in Figs. 1(a)-(d) are shown in Figure 1(e) as well as a line profile from the first AFM scan 30 mins after exfoliation. The height across the flake more than doubles over the measurement period. Using the AFM scans, the total volume of the flake and absorbed water is recorded with time and shown in Figure 1(f). The volume increases rapidly in the first 15 hours at a rate of $\approx7$ $\mu m^3$/min at which point it stays relatively constant as larger droplets that completely cover the surface start to coalesce to form one large bubble. At 60 hours the volume again starts to increase but at a slower rate of $\approx2$ $\mu m^3$/min. After five days the volume has increased by more than 200\%. 

\begin{figure}
\centerline{\includegraphics[width=5in]{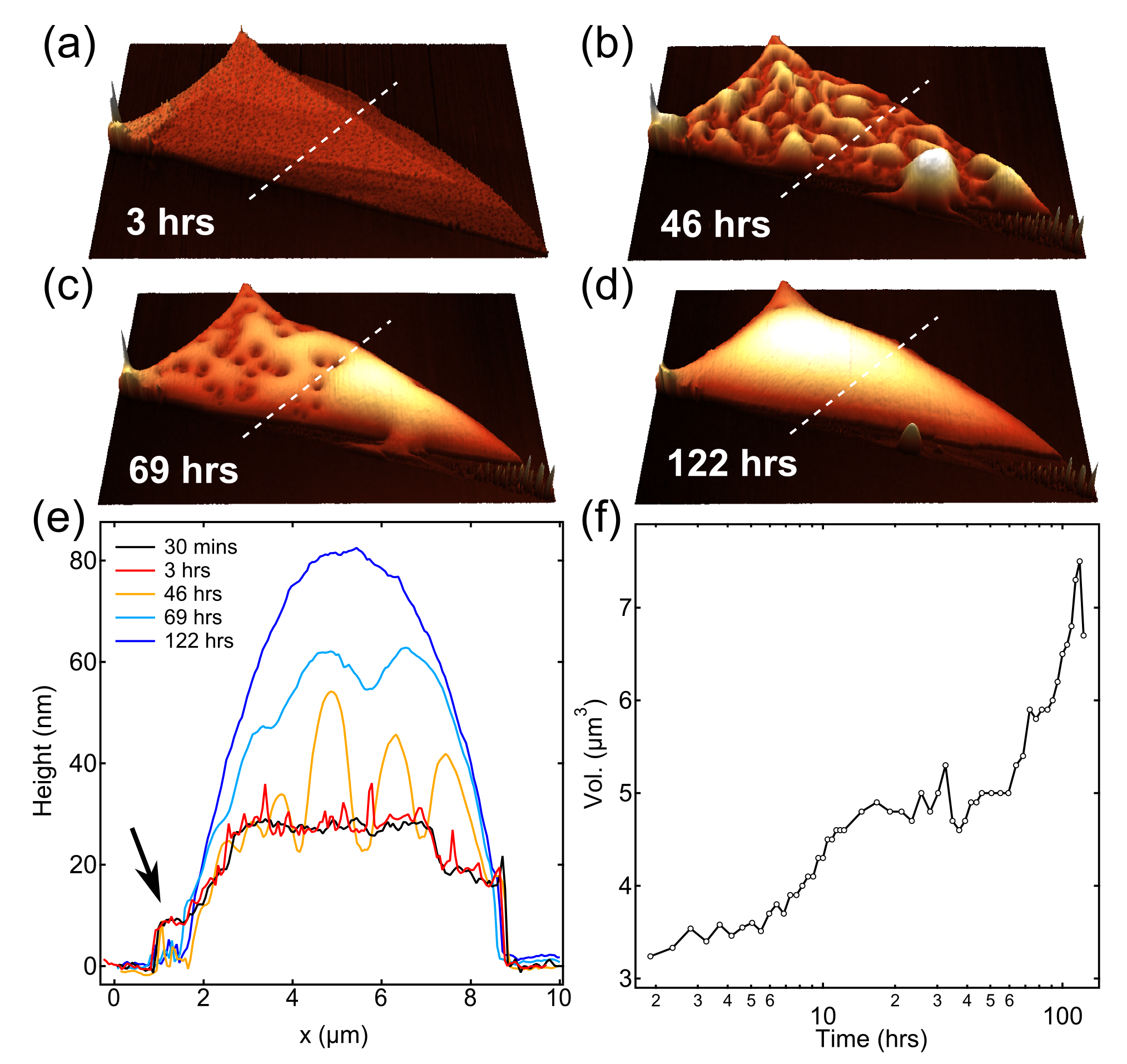}}
\caption{\label{} (a)-(d) Selected AFM scans of a BP flake in air taken at (a) 3 hours, (b) 46 hours, (c) 69 hours, and (d) 122 hours after exfoliation. Dotted white lines indicate the position of the line profiles shown in (e). (e) Line profiles taken across the BP flake from the scans in (a)-(d) as well as a line profile for the AFM scan taken directly after exfoliation. (f) Total volume of the flake and water over the measurement period.}
\end{figure}

After continuous AFM scanning, the samples are subsequently put in vacuum ($~10^{-6}$ mbar) and heated ($>100 ^\circ$C) to remove absorbed water. From optical images and AFM scans after exposure and water removal, we find that the flakes have changed considerably from their exfoliated form. The optical images (see Supplemental Information) and AFM images in Figure 2 before and after exposure for the same flake in Figure 1 reveal an apparent removal of the material from the thinnest part of the flake as well as the presence of left-over residues (white bubble in panel (b)). The thinning of the flake can also be seen in the line profiles in Figure 1(e) at the location of the arrow. The thinnest part of the flake (8 nm) nearly disappears after 46 hours of exposure. Full line profiles from AFM scans before and after exposure, shown in Figure 2(c), also show the greatest degradation for the thinner part of the flake. The overall reduction in thickness suggests that exposure to ambient conditions results in a sort of layer-by-layer thinning of the flakes from the top down. 

To further highlight this thinning effect, Figures 2(d)-(f) show optical and AFM images of another BP flake on the same substrate and subjected to the same ambient exposure as the flake in Figures 2(a)-(b). Color contrast difference of the optical images of the flake (78\% before exposure, 34\% after) indicate a drastic reduction in thickness as well as a reduction in flake area (marked by the white dotted line in panel (e)). An AFM image of the flake in Figure 2(f) reveals that the original area is preserved but the edges have been reduced to single layer thicknesses ($\approx 0.7$ nm). Residues are again found on this flake and they can be seen as white saturated regions in the AFM image in Figure 2(f). The phase image from the AFM scan in Figure 2(f) (see Supplemental Information) shows a clear distinction between the BP flake, the left-over residues, and the SiO$_2$ substrate, all having different material densities, suggesting that the last layer of the flake is still BP but further investigation is required \cite{magonov97}. 

Recent DFT calculations from Ziletti et al. suggest that oxygen (under light illumination) could be responsible for crystal degradation \cite{ziletti14}. At large enough concentrations, oxygen that has penetrated into the lattice, overcoming an activation energy of 0.69 eV with the help of light, can create stress in the crystal lattice and break it apart forming phosphorus trioxide or phosphorus pentoxide. This is supported by a complimentary study on the degradation of black phosphorus that suggests a photo-induced oxidation reaction in the presence of oxygen absorbed in water \cite{favron}. These studies agree with the degradation and thinning we observe here for flakes that take on water in ambient light conditions. Furthermore, we note that the thinner flakes absorb water faster than the thicker ones (see video clip and additional flakes in the Supplemental Information for further details) thereby resulting in a thickness dependent wettability which enhances the degradation for thinner flakes, this, also reported in the complimentary study \cite{favron}. Finally, it must also be noted that while exfoliated flakes are free from extended defects, point defects, commonly observed in 2D materials\cite{hashimoto04, zhou13}, could play a substantial role in the degradation process and require further study. By controlling atmospheric conditions, one could envision control over the thickness of BP flakes using this observed layer-by-layer thinning effect. 

\begin{figure}
\centerline{\includegraphics[width=3in]{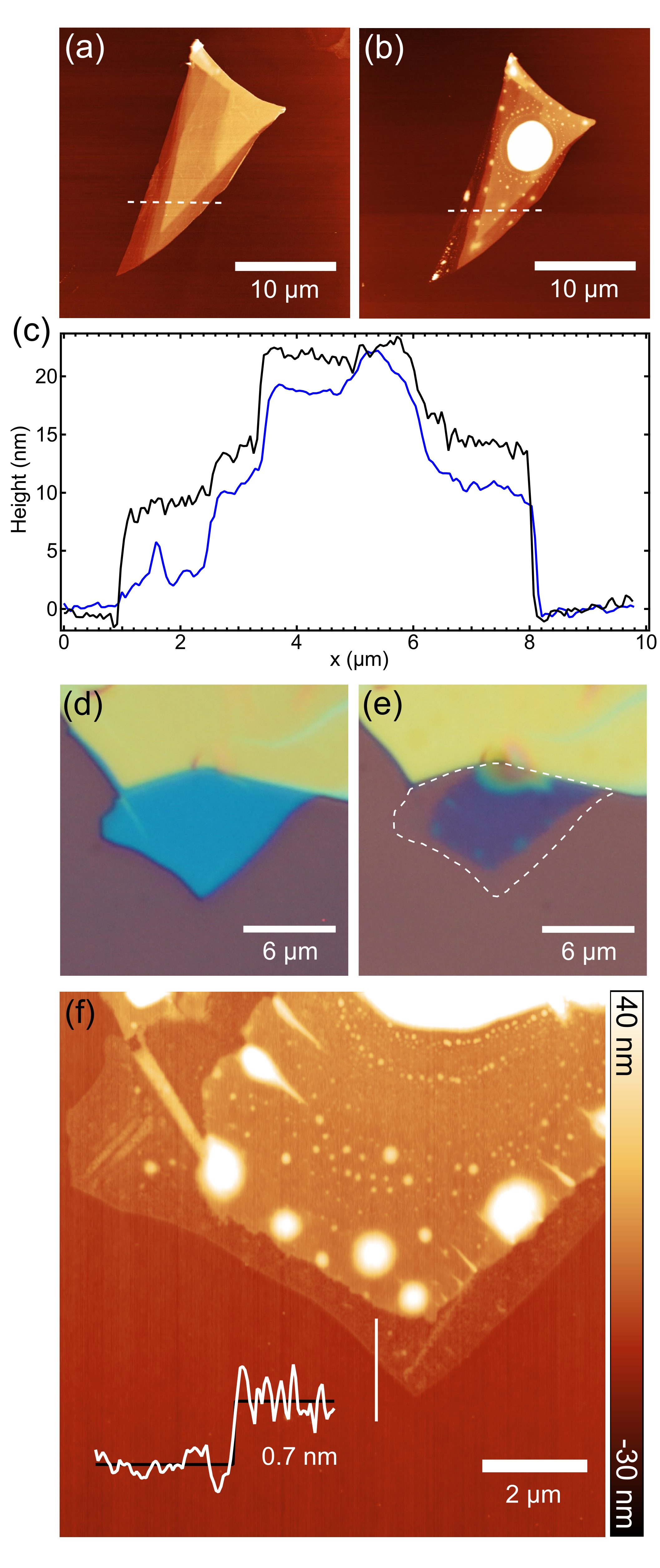}}
\caption{\label{} (a). AFM image taken approximately 30 mins after exfoliation for the same flake shown in Figure 1. (b) AFM image of the same flake in (a) after exposure and pumping/heating. The white saturated region shows the left-over residues. White dotted lines show the location of the line profile shown in (c). Note: scale for AFM scans is -10 nm (black) to 45 nm (white). (c) AFM height scans taken at the locations of the dotted lines in (a) and (b). Black curve shows the height before exposure and the blue curve after. (d). Optical image for a second flake on the same substrate taken directly after exfoliation and before 5 days of ambient exposure. (e) Same flake in (d) after exposure and pumping/heating. White dotted line marks the edge of the original flake. (f) AFM topography scan of the flake shown in panel (e). The white line marks the location of the line profile shown in the inset (white curve).}
\end{figure}

In the following, we discuss the the effect of the unique hydrophilic behavior of BP on the electrical performance of FETs. Few-layer BP FETs are fabricated on Si/SiO$_{2}$ substrates with pre-patterned electrodes (the SiO$_2$ is used as a gate dielectric and the heavily doped Si substrate is used as a back gate electrode). Source and drain electrodes are created using a shadow mask and deposition of 30 nm of Au with a 5 nm Ti adhesion layer. Few-layer BP flakes are then transferred onto the pre-patterned electrodes (contact is made to the top of the gold electrodes) using a recently developed deterministic transfer method outlined in Ref. ~\citenum{castellanos14}. Transferring BP flakes to pre-patterned electrodes considerably reduces the time that the flakes are exposed to air between exfoliation and characterization. Typical devices are fabricated and put in vacuum 10 minutes after flake transfer ensuring the measurement of nearly pristine flakes. Twelve devices were fabricated and measured in ambient conditions and we focus on one device in the main text for consistency.  

\begin{figure}
\centerline{\includegraphics[width=2.8in]{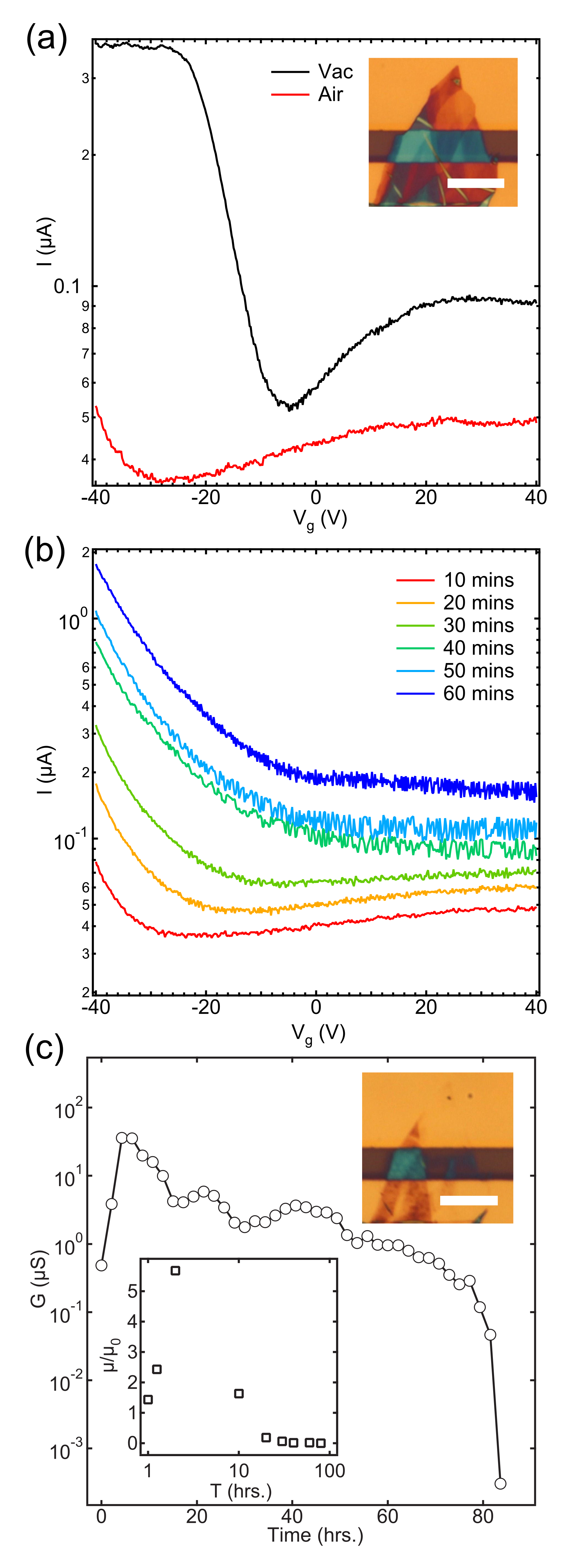}}
\caption{\label{} (a) Transfer characteristics ($I-V_g$) of the FET device in vacuum before exposure (black curve) and directly after breaking the vacuum (red curve) taken at $V_b$ = 100 mV. Inset shows an optical image of a BP flake before longterm ambient exposure (scale bar is 10 $\mu$m). (b) Transfer characteristics ($I-V_g$) at selected times over the first hour of exposure (curves are offset by 100 nA for clarity). (c) Conductance vs. time after long term exposure to air taken at $V_b$ = 100 mV, $V_g$ = 40 V. Upper inset shows an optical image of the device after measurement (scale bar is 10 $\mu$m). Lower inset shows the two terminal estimated hole mobility (scaled by the initial vacuum hole mobility) vs. time. }
\end{figure}

After deterministic transfer of a few-layer BP flake, FETs are characterized in vacuum and at room temperature. Figure 3(a) shows the two terminal transfer characteristics for a representative device in vacuum at a bias voltage of 100 mV (black curve). The device displays ambipolar transport with current saturation and estimated hole (electron) mobilities of 6.5 cm$^2$/Vs (1.2 cm$^2$/Vs) calculated from the measured transfer curve and estimated insulator capacitance using a parallel plate capacitance model \cite{horowitz99}. Note that we determine the mobilities from two terminal measurements which are known to underestimate the BP FET mobilities by about a 50\% \cite{liu14}. The mobility value of this device is slightly lower than that expected for BP devices with a comparable thickness (5 to 100 cm$^2$/Vs, according to Ref. \citenum{li14}). This might be due to a higher contact resistance between the flake and the top gold electrode due to the fabrication process, based on directly stamping onto pre-patterned electrodes. However, while we sacrifice interface contact for stamped flakes, we believe stamp transfer is the most efficient method to study pristine BP flakes that have had minimal exposure to the environment.

Once exposed to air, there is an instant shift of the threshold voltage to the left and an overall decrease in conductance (red curve in Figure 3(a)). We find that this short timescale effect is reversible over several cycles (see vacuum to air cycling for another device in the Supplemental Information), indicating that it is caused by physisorption of atmospheric species such as oxygen and nitrogen before the condensation of water. To investigate further this effect, BP FETs were measured separately in oxygen and nitrogen environments (see Supplemental Information for details). We find that the initial effect is a combination of the physisorption of oxygen which broadens and decreases the OFF state current and the physisorption of nitrogen which shifts the threshold voltage to the left. These two effects are responsible for the shift in threshold voltage and decrease in conductance shown in Figure 3(a) (red curve). As the BP flake starts to absorb water, a transition occurs from the short-term physisorption effects to strong p-type doping (Figure 4(b)). This evolution occurs over the first hour of continuous measurement. Figure 3(c) shows the conductance vs. time after the crossover and for the further collection of water and final breakdown (see Supplemental Information for transfer curves). The conductance steadily increases by two orders of magnitude over 5 hours after which degradation starts to take over and the conductance begins to drop. The transfer curves become almost featureless and the current through the device reaches zero after $\approx80$ hours in air. An optical image of the device after measurement is shown in the inset of Figure 3(c). The material on the electrodes has been etched away as well as some of the material in the channel. We note that the degradation appears to be more severe on the electrodes and could be due to the higher catalytic behavior of gold as compared to SiO$_2$. This violent degradation is in contrast to FETs made with other 2D materials such as graphene and MoS$_2$ which show comparatively subtle changes in ambient conditions \cite{shin11, qiu12}. Preliminary studies on top-sided encapsulation of BP FET devices using boron nitride (BN) to preserve the BP flake proved unsuccessful (see Supplemental Information). We suspect that water and oxygen enter at the BN-SiO$_2$ and BN-Au interfaces leading to eventual breakdown. Full encapsulation between two layers of BN, as shown for graphene, could be a method to preserve BP FET devices \cite{wang12}. We found, however, that single BP flakes could be preserved over 1 week simply by sandwiching them between two layers of PDMS (see Supplemental Information). 

We have studied the environmental stability of few-layer BP flakes through two methods. Isolated BP flakes were exfoliated and continuously scanned using an AFM in tapping mode. We find that BP has a strong affinity for water and observe an increase of over 200\% in volume due to water absorption over several days. In addition, after longterm exposure to air, we observe a layer-by-layer etching of the BP flakes which can be potentially used to obtain single layer BP. Finally, few-layer BP FETs were fabricated and the FET characteristics were measured in vacuum and air. These devices show a significant effect of the environment on their FET characteristics, initially from the physisorption of oxygen and nitrogen (which takes place on a timescale of minutes), and subsequently from the absorption of water leading to degradation and breakdown (a slower process that takes several hours to days). 

This work was supported by the European Union (FP7) through the program RODIN, the Dutch organization for Fundamental Research on Matter (FOM), and NWO/OCW. A.C-G. acknowledges financial support through the FP7-Marie Curie Project PIEF-GA-2011-300802 (‘STRENGTHNANO’).

\bibliography{bib}

\end{document}